\documentclass[%
 reprint,
 amsmath,amssymb,
 aps,
floatfix,
]{revtex4-2}

\usepackage{xcolor}
\usepackage{graphicx}% Include figure files
\usepackage{dcolumn}% Align table columns on decimal point
\usepackage{bm}% bold math
\begin{document}

\preprint{APS/123-QED}

\title{Spontaneous symmetry breaking and the dynamics of three interacting nonlinear optical resonators with gain and loss}
% Force line breaks with \\

\author{D. Dolinina}
\email{d.dolinina@metalab.ifmo.ru}
\author{A. Yulin}%
\email{a.v.yulin@corp.ifmo.ru}
\affiliation{Faculty of Physics, ITMO University, Saint Petersburg 197101, Russia}%

\date{\today}% It is always \today, today,

\begin{abstract}
The dynamics of two active nonlinear resonators coupled to a linear resonator is studied theoretically. Possible stationary states and its dynamical stability are considered in detail. The spontaneous symmetry breaking is found and it is shown that this bifurcation results in the formation of asymmetric states. It is also found that the oscillating states can occur in the system in a certain range of parameters. The results of the analysis of the stationary states are confirmed by direct numerical simulations. The possibility of the switching between different states is also demonstrated by numerical experiments.
\end{abstract}

\maketitle

\section{Introduction}

It is well known that a symmetric system always have a symmetric solution. However this solution must not necessarily be dynamically stable and it can happen that a symmetric system shows a stable asymmetric solution. In some system the symmetric system looses its stability and the switching to an asymmetric state happen when one of the parameters exceeds a threshold value. This is known as spontaneous symmetry breaking (SSB). Such phenomena is one of the most fundamental processes in nonlinear science, that is why it is under great attention in many fields of physics \cite{Salam_theory,Higgs,Zibold,Jiaming,Malomed_book} for decades. In particular, SSB appears in many nonlinear optical systems, such as waveguide arrays \cite{Malomed1,Malomed2}, double-well systems \cite{Matuszewski,Mayteevarunyoo, chen}, grated waveguides \cite{Krasikov2018,Dolinina1,Dolinina2} and dual-core fibers \cite{albuch,birnbaum}.

Besides, this effect was found in systems of two- or three- coupled nonlinear waveguides \cite{schmidt1992,bernstein1992,molina1992,molina1993,eilbeck1995,deering1993,menezes2007}. It was shown that the systems of three interacting identical nonlinear couplers \cite{schmidt1992,bernstein1992} or of two interacting nonlinear and one linear couplers \cite{molina1992,molina1993,eilbeck1995} provides multistability caused by the appearance of asymmetric states. These systems are especially interesting from the perspective of all-optical device switching. The problem of two nonlinear conservative couplers interacting with a linear conservative coupler is considered in \cite{molina1993} where different propagation regimes, including chaotic ones, are reported.

%it is demonstrated that the configuration of two nonlinear and one linear coupler have switching properties superior to that of two and three all nonlinear coupler configurations \textcolor{red}{What does it mean?}.

In our work we consider the similar system consisting of two nonlinear resonators with linear gain (micro-lasers) saturated by nonlinear ( cubic ) losses. We also assume that these active resonators have conservative nonlinearity which makes their resonant frequency dependent of the intensity of the field inside the resonators.  The active resonators do not interact to each other directly but both of them coupled to another resonators situated between them. This resonator is linear with some losses and the resonant frequency detuned from the resonant frequency of the active resonators in the linear regime. The coupling to this linear resonator introduced the effective coupling between the active resonators. These complex of three resonators we further refere as trimers. In the present work we focus on the stationary states and the symmetry breaking that occur in these trimers. We believe that the results reported in the paper is of interest not only from the point of view of applied mathematics but can be used to design switchable multi-stable sources of coherent light or be used for all-optical calculations. 

Let us remark that the conservative analogue of this trimer is considered in \cite{eilbeck1995} where the symmetry breaking is reported. However, the presence of the dissipative terms affects the bifurcation strongly and, what can be even more important, makes some states to be attractors and thus allow for the switching between different stationary states. So the problem of the formation and the switching between the states becomes of interest. For example, below we show that the instability of time-independent states can result in the switching of the trimers to the states where the intensities of the fields in the resonators experience periodic oscillations. 

Let us briefly discuss possible physical realizations of such systems. We believe that this can be achieved in the systems of interacting microlasers interacting because of the non-perfect localization of the field inside the resonators. Thus the resonators separated by a small gaps can interact through the evanescent field of their eigenmodes.  One of the promising allowing for fabrication of such devices is perovskites that are capable to provide large optical gain and thus obtain lasing in dielectric resonators of small volumes and relatively large radiative losses  \cite{veldhuis2016perovskite, sutherland2016perovskite,wang2018recent}.

Another physical realization of the suggested system of the oscillators is exciton-polariton systems where the polariton condensation occurs in the interacting micropillars. These systems are realized experimentally and have been studied for more than a decade \cite{Bajoni2008,Ferrier2011,klein2015,schneider2016,amo2016exciton,kalinin2020polaritonic}.

It is also worth mentioning that the effect known as Bound State in the Continuum that have been actively studied in resent time in optical systems including nonlinear ones  \cite{hsu2013observation, hsu2016bound, koshelev2020subwavelength}. Indeed, in the case of BIC the radiation losses disappear because of destructive interference that eliminate the radiation field completely. In the proposed system the coupling to the linear resonators introduces some additional losses for the active nonlinear resonators. The important fact is the effective losses seen by the active resonators depend on the mutual phase of the field in the resonators. Indeed, one can easily see that if the mutual phase is equal to $\pi$ the total driving force for the middle resonator is equal to zero. So the passive linear oscillator is not excited and, consequently, it does not contribute to the effective losses experienced by the active resonators. This is similar to what happens in BIC systems considered in \cite{bulgakov2011symmetry}.   

% Thus, there are reasons to believe that modern technologies paves a way to manufacture the system of interacting lasers where the effects discussed in the paper can possibly be observed. 

The paper is organized as follows. In the next Section we discuss the physical system under consideration and introduce a mathematical model describing the dynamics of the optical fields in the resonators in terms of slow varying amplitudes. In the Sections III-V we do comprehensive analysis of the anti-symmetric, symmetric and asymmetric stationary states and bifurcations taking place in the active trimers. In Sec. VI we perform a numerical experiment demonstrating the formation of states considered in Sec. III-V. Finally in the the Conclusion we briefly summarize the main findings reported in the paper.

\section{The physical system and its mathematical model}

\begin{figure}[t]
\centering
\includegraphics[width=\linewidth]{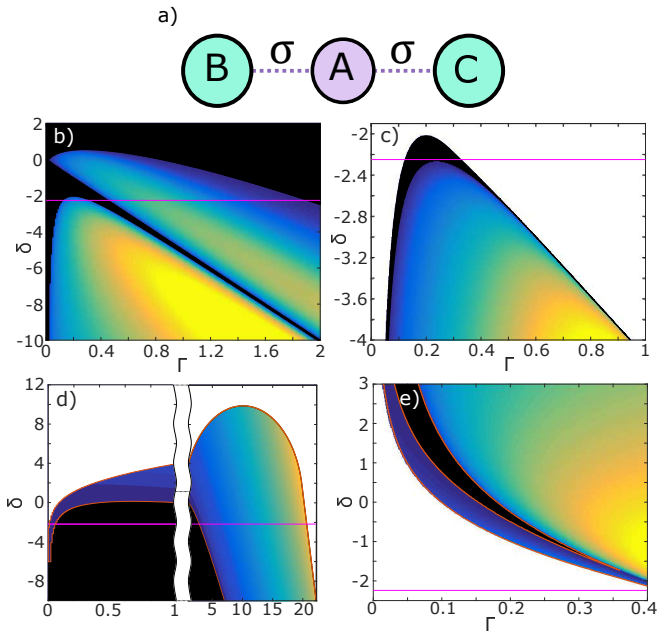}
\caption{a) The sketch of the considered system; The growth rate of the unstable perturbations on the parameter plane of the detuning $\delta$ and gain $\Gamma$ for (b) antisymmetric, (c) hybrid, (d) symmetric I and (e) symmetric II states. Black color corresponds to dynamically stable states (max(Re$(\lambda)$)$\leq$ $0$). Region of parameters where chosen state does not exist is shown by white color. The pink line corresponds to detuning used for the calculation of the bifurcation diagram shown in Fig.\ref{fig1}.}
\label{fig2}
\end{figure}

Let us start with the discussion of the physical system in question. We consider a system consisting of three interacting resonators schematically shown in Fig.\ref{fig2}(a). The right and the left resonators have linear gain saturated by nonlinear losses. From the experimental point of view each of these resonator can be seen as nano-lasers pumped above the threshold. These resonators also have conservative Kerr nonlinearity so that their resonance frequencies depend on the intensity of the field inside the resonators. In this paper we consider the case when the left and the right resonators are identical.  Thus, we consider only symmetric trimers.  The right and the left resonators do not interact directly, but they both coupled to the middle resonator which is passive, linear and has resonant frequency detuned from the resonant frequency of the other two resonators of the trimer.

To describe the dynamics of the trimer we use slow varying amplitudes approximation characterizing the field in each of the resonator by a complex amplitude of the resonator eigenmode. Then the equations can be written as follows:
\begin{subequations}
	\label{eq:trimer}
	
	\begin{equation}
\partial_t B = \Gamma B - \beta |B|^2 B + i \alpha |B|^2 B + i \delta B + i \sigma A,
	\label{tr:1}
	\end{equation}
	
	\begin{equation}
\partial_t C = \Gamma C - \beta |C|^2 C + i \alpha |C|^2 C + i \delta C + i \sigma A ,
	\label{tr:2}
	\end{equation}
	
	\begin{equation}
\partial_t A = -\gamma A + i \sigma (B + C),
	\label{tr:3}
	\end{equation}
	\end{subequations}
where $B$ is the amplitude of the left resonator, $C$ is the amplitude of the right resonator, $A$ is the amplitude of the middle resonator, $\gamma$ is the losses in the middle resonator,  $\Gamma$ is linear gain of the left and the right resonators, $\beta$ characterizes the strength of the nonlinear losses and $\alpha$ is the coefficient of the conservative nonlinearity in active resonators, $\delta$ is the detuning of linear resonance frequencies of the left and the right resonators from the frequency of the middle resonator, $\sigma$ is the coupling strength of the middle resonator to its neighbours.

Let us acknowledge that the excitation of the linear resonator ``A'' depends on the mutual phase of the oscillations in the resonators ``B'' and ``C''. One can easily see that there is a mode $B=-C, \quad A=0$ such that the middle resonator is not excited. It is obvious that there must exist also a mode $B=C, \quad A \neq 0$. It is instructive to rewrite the equations in the form of the ``symmetric'' $U_s = \frac{B + C}{\sqrt{2}}$ and an ``antisymmetric'' $U_a = \frac{B - C}{\sqrt{2}}$ modes. In new variables the system of equations for the trimer reads:

\begin{subequations}
	\label{eq:sym_antisym}
	
	\begin{equation}
\partial_t U_s = (\Gamma + i \delta) U_s + (i \alpha - \beta) (K_1 U_s + M U_a) + i \sqrt{2}\sigma A,
	\label{eq:sym}
	\end{equation}
	
	\begin{equation}
\partial_t U_a = (\Gamma + i \delta) U_a + (i \alpha - \beta) (K_2 U_a + M^* U_s),
	\label{eq:antisym}
	\end{equation}
	
	\begin{equation}
\partial_t A = -\gamma A + i \sqrt{2} \sigma U_s.
	\label{eq:A_}
	\end{equation}
\end{subequations}

where $K_1 =  \dfrac{1}{2}(|U_s|^2 + 2 |U_a|^2)$, $K_2 = \dfrac{1}{2}(2 |U_s|^2 + |U_a|^2)$ and $M = \dfrac{1}{2}U_s^*U_a$.

From these equations it is easy to conclude that in the linear regime there are one anti-symmetric mode and two symmetric modes with different eigenfrequencies $\omega_a=\delta-i\Gamma $ and $\omega_{s\pm}= \frac{1}{2}\left(  \delta+i(\gamma-\Gamma) \pm i\sqrt{(\gamma+\Gamma+i\delta)^2 -8\sigma^2}\right)$ correspondingly. It is important for us that the losses of the anti-symmetric mode is always lower than the losses of at least one of the symmetric mode. This is the consequence of the symmetry of the system which provides that in the anti-symmetric mode the middle resonator is not excited at all. 

With the increase of $\Gamma$ the effective gain of the antisymmetric mode changes its sign and becomes positive. Simple algebra shows that at this $\Gamma$ the effective gain for both symmetric modes is negative.
So the anti-symmetric mode exceeds the lasing threshold first. At higher gain $\Gamma$ the symmetric mode also starts growing. Because of the nonlinearity the growth of the modes is saturated and a stationary state with constant amplitude can form.

The stationary symmetric nonlinear solutions can be found in the form $\vec W_s = (U_s \neq 0, U_a = 0, A \neq 0 )^T$ and $\vec W_a= (U_s = 0, U_a \neq 0, A = 0)^T$. Let us note that these modes have the same symmetry as the linear eigenmodes. However, in the nonlinear regime there may exist the stationary states with all non-zero components $U_s \neq 0$, $U_a \neq 0$, $A \neq 0$. We will refer to these states as hybrid nonlinear states meaning that they can be seen as a sum of a symmetric and anti-symmetric components. The next Sections are devoted to the detailed investigation of the nonlinear states.

\section{The  anti-symmetric states of the trimers}

Now let us consider nonlinear stationary states in more details. We search these states in a form $A = A_0 e^{i\omega t}$, $B = B_0 e^{i\omega t}$ and $C = C_0 e^{i\omega t}$, where $A_0$, $B_0$ and $C_0$ are unknown complex amplitudes and $\omega$ is unknown frequency. The important fact is that the equations (\ref{tr:1})-(\ref{tr:3}) are invariant in respect to the transformation $\alpha \rightarrow -\alpha$, $\delta \rightarrow -\delta$, $A \rightarrow - A^{*}$, $B \rightarrow B^{*}$, $C \rightarrow C^{*}$ and $\omega = - \omega$. 
This means that without loss of generality we can restrict our consideration to either positive or negative sign of Kerr nonlinearity coefficient $\alpha$. 
We choose to set $\alpha$ to be positive. Let us also mention that the coupling coefficient  $\sigma$ can also be chosen to be positive without loss of generality  because of the invariance of the equations in respect to the transform $\sigma \rightarrow - \sigma $,  $A \rightarrow - A $.  

Let us start with consideration of the antisymmetric states. The non-trivial states of this symmetry  exist for $\Gamma>0$. The amplitude  and frequency of antisymmetric states can be found analytically $|B_0|=|C_0|=\sqrt{\Gamma/\beta}$ and $\omega = \alpha \Gamma/\beta + \delta$. We found the areas of existence and analysed the stability of the state numerically. Our results is summarized in Fig.\ref{fig2}(b) where the instability growth rate of the anti-symmetric state is shown as a function of the gain $\Gamma$ and the detuning $\delta$, the area of the existence of the stable states is shown by black color.

\begin{figure}[bt]
\centering
\includegraphics[width=\linewidth]{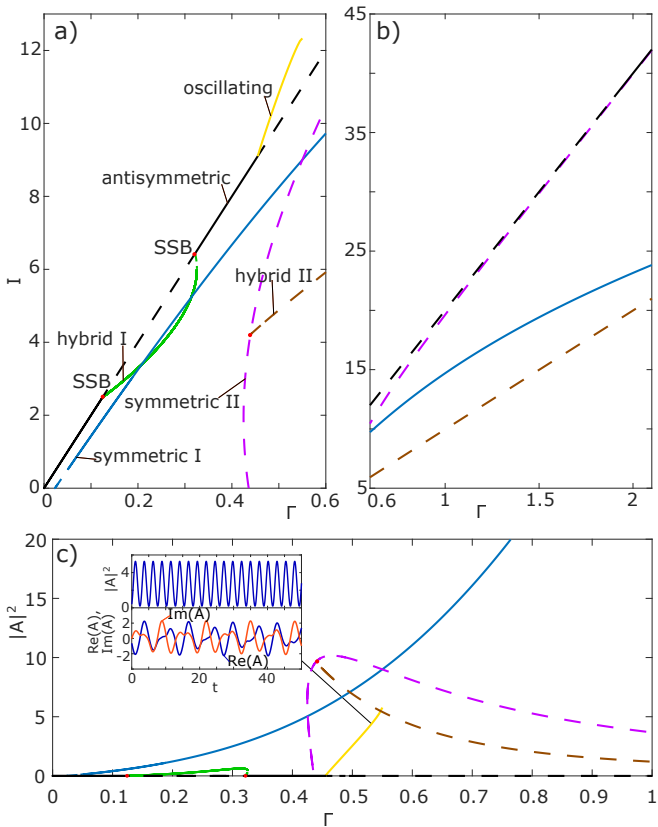}
\caption{Bifurcation diagram showing the dependence of field intensity in (a)-(b) active resonators $I = |B|^2 + |C|^2$ and in (c) passive resonator  $|A|^2$ of time-independent stationary states on the gain $\Gamma$. Black line shows the intensity dependence for antisymmetric states; the blue and magenta curves are for the symmetric states; green and red lines shows the bifurcation curve for the hybrid states of two kinds. We use dashed lines for the dynamically unstable states and solid lines for the dynamically stable ones. The yellow curves shows the maximum amplitude of the dynamical state with the intensity periodically oscillating in time. The red dots mark the spontaneous symmetry bifurcation. Parameters are: $\alpha = 0.5$, $\delta = -2.25$, $\gamma = 0.1$, $\sigma = 1$, $\beta = 0.1$ The inset shows bifurcation curves of hybrid states with $\delta = -2.1$ and $\delta = -2.4$.}
\label{fig1}
\end{figure}

One can see that for  sufficiently large positive detunings the state is stable for all values of gain $\Gamma$. However, for lower detuning there may exist one or two regions of instability. The bifurcation diagram for the stationary anti-symmetric states $\vec W_a$ showing the dependence of the field intensity in both laser elements $I = |B|^2 + |C|^2$ on gain $\Gamma$ is shown in Fig.\ref{fig1}(a-b) for $\delta=-2.25$ by black color. On the same figure the bifurcation diagrams of the states of the other kinds are shown.

\begin{figure}[bt]
\centering
\includegraphics[width=\linewidth]{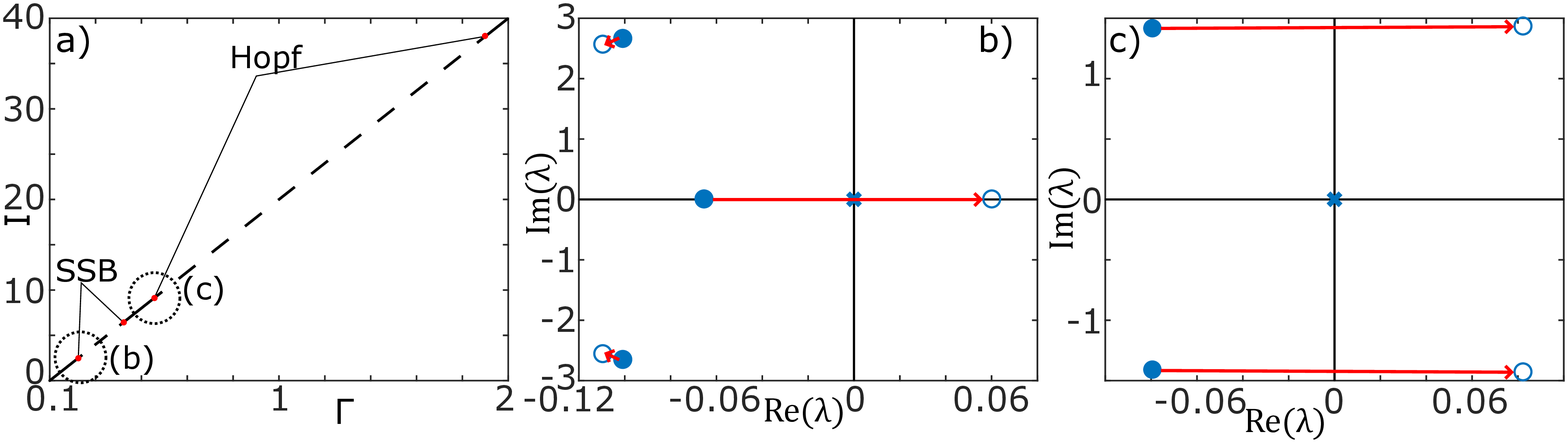}
\caption{(a) Bifurcation diagram of the antisymmetric state showing the stability changes of the state. Four eigenvalues with the larges real parts are shown in panel (b) just before (solid circules) and after (open circles) the pitchfork SSB bifurcation. The motion of the eigenvalues for the Hopf bifurcation is shown in panel (c).  Red arrows show the motion of eigenvalues during the bifurcation. Zero eigenvalues associated with phase symmetry of (\ref{eq:trimer}) are marked by a blue cross in both panels (b) and (c).}
\label{anti_bifs}
\end{figure}

We studied the stability of the antisymmetric state (Fig.\ref{anti_bifs}(a)) and it is found that for negative $\delta$ of sufficiently large absolute values both spontaneous symmetry breaking and Hopf bifurcations take place. First, with the increase of $\Gamma$ the anti-symmetric states become unstable against the perturbations having the structure of the symmetric mode. The motion of the eigenvalues governing the behaviour of the weak perturbations of the anti-symmetric state is shown in Fig.\ref{anti_bifs}(b). So we can conclude that the symmetry breaking of the anti-symmetric state goes through a supercritical pitchfork bifurcation.  

As a result of such bifurcation the hybrid states with non-zero components of both symmetric and antisymmetric modes appear. The bifurcation curve for these states is shown in Fig.\ref{fig1} by green color. The hybrid states are characterized by non-equal field amplitudes of the resonators $|B| \neq |C|$ and nonzero amplitude of the middle resonator  $A \neq 0$, see Fig.\ref{fig1}(c), showing field intensity $|A|^2$ in the middle resonator as a function of the gain. We refer to these asymmetric states as hybrid-I states and will consider them in more detail below.

The analysis shows that at a higher gain the anti-symmetric state restores its stability colliding with the hybrid-I states in another pitchfork bifurcation. Let us mention that depending on the detuning this bifurcation can be either super- or subcritical, see Fig.\ref{hyb_bifs}. 

With the further increase of the gain the anti-symmetric state  gets destabilized through a supercritical Hopf bifurcation which gives birth to an oscillatory state. The motion of the eigenvalues for this case is shown in Fig.\ref{anti_bifs}(c). Finally, at even higher gain the anti-symmetric state restores its stability again through a subcritical Hopf bifurcation, see Fig.\ref{anti_bifs}(a). 

Let us note that the oscillatory state bifurcating from the anti-symmetric one is quasi-periodic, see the inset of Fig.\ref{fig1}(c) illustrating the temporal dynamics of the intensity of the field and the temporal evolution of the real and imaginary parts of the field. Let us also mention here, that at some threshold linear gain the dynamical state becomes unstable and the instability switches the system into the symmetric state, which will be considered below. The bifurcation curve of the oscillatory state showing the dependence of the maximum intensity of the oscillations on the linear gain $\Gamma$ is shown in Fig.\ref{fig1} by the yellow curve within the range of dynamical stability of the state. 

The stability of the anti-symmetric states becomes different for the detunings $\delta>\delta_{cr1} \approx -2.02$. At $\delta>\delta_{cr1}$ no hybrid states bifurcate from the anti-symmetric state. When the detuning exceeds another threshold value $\delta_{cr2}\approx 0.56$ two Hopf bifurcation merge and the anti-symmetric state becomes stable for all gain $\Gamma$.

 \section{Hybrid states}
 
Now let us return to the hybrid states. It is worth noting here that the hybrid states appearing as a result of spontaneous symmetry bifurcation are double degenerated. In other words, if  $B=B_0$, $C = C_0$, $A = A_0$ and $\omega = \omega_0$ is a solution then  $B=C_0$, $C = B_0$, $A = A_0$ is also a solution having the same frequency $\omega_0$. The area of the existence and the stability of these hybrid-I states are shown in Fig.\ref{fig2}(a).

\begin{figure}[bt]
\centering
\includegraphics[width=\linewidth]{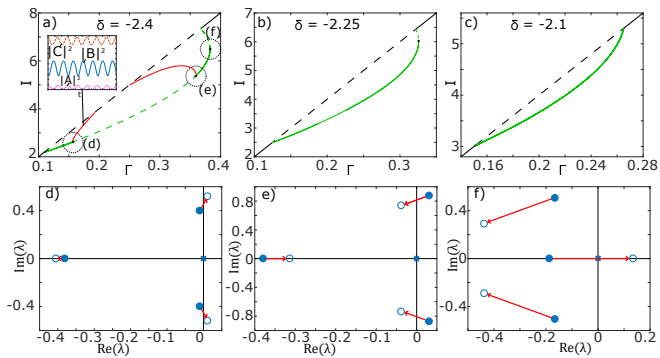}
\caption{ Panels (a), (b) and (c) show the bifurcation diagrams of the hybrid states for different detunings $\delta = -2.4$, $\delta = -2.25$ and $\delta = -2.1$. Panels (d)-(f) illustrate the motion of the eigenvalues with largest real parts governing the dynamics of the weak excitations on the hybrid state shown in (a). The solid circles correspond to the eigenvalues before and the open circles to the eigenvalues after the bifurcations. The red arrow show the directions of eigenvalues motions. The yellow curves in (a) show the maximum amplitudes of dynamically stable limit cycles bifurcated from the hybrid stationary state. The inset in (a) shows temporal dependencies of the field intensities for the case of the oscillating regime.}
\label{hyb_bifs}
\end{figure}

First, we consider the bifurcations of the hybrid states for large negative detunings $\delta  < -2.26$. As it is said above the hybrid state branching off the anti-symmetric state is stable in the vicinity of the bifurcation point.  Then, at a threshold gain the hybrid state looses its stability via a supercritical Hopf bifurcation, see Fig.~\ref{hyb_bifs}(a) and  Fig.~\ref{hyb_bifs}(d) where the motion of the eigenvalues with the largest real part are shown.  As a result of the bifurcation a stable limit cycle appears. The bifurcation curve for this oscillatory state is shown in Fig.~\ref{hyb_bifs}(a) by the yellow line, the dynamics of the field intensities in all three resonators is shown in the inset of Fig.~\ref{hyb_bifs}(a).

 The stability of the hybrid state is restored at a higher  gain through another supercritical Hopf bifurcation (the eigenvalues motion is shown in Fig.~\ref{hyb_bifs}(e) ). Then the hybrid states undergoes fold bifurcation and becomes unstable (the eigenvalues motion is shown in Fig.~\ref{hyb_bifs}(f) ). Finally, the hybrid states merges with the anti-symmetric states via a subcritical pitchfork bifurcation.   
 
 With the decrease of the absolute value of the negative detuning Hopf bifurcations of the hybrid-I states collide and disappear at $\delta_{cr3}\approx-2.26$. However, the fold bifurcation of the states survives until $\delta_{cr4} \approx -2.14$. The bifurcation curve of the hybrid state for this case is shown in Fig.~\ref{hyb_bifs}(b). For higher detunings $\delta>\delta_{cr4}$ the fold bifurcation disappears and the hybrid states merge with the antisymmetric state through a supercritical pitchfork bifurcation, see Fig.~\ref{hyb_bifs}(c). As one can see, in this case the hybrid states are stable within the whole range of existence. Finally, the hybrid states cease to exist for $\delta >\delta_{cr1}$.

% At even higher linear gain $\Gamma$ the anti-symmetric state regains the dynamically stable again.  Let as mention that 

% to the lower values of the gain and the width of the instability range shrinks. At some threshold value the instability disappears and below this threshold the antisymmetric state is dynamically stable for all gain values, see Fig.\ref{fig2}(b) where the bifurcation curve is shown for $\delta > 0.556$.

%($\delta>-2.25$) makes the the instability region wider and shifts it to higher gain values, see Fig.\ref{fig2}(b). On contrary, the smaller 
%, while with decreasing ($\delta>-2.25$) of detuning the unstable region becomes smaller and moves to lower gain, see Fig.\ref{fig2}(b). 
%From the figure it is clearly seen that for $\delta > 0.556$ the antisymmetric state becomes dynamically stable for all gain values.

\section{Symmetric states}

Now let us consider the symmetric states $B=C$, $A\neq 0$. There may be two kinds of these solutions bifurcating from the trivial state. The bifurcation curves for these states are shown in Fig.\ref{fig1} in blue and magenta colors, correspondingly. The ``symmetric-I'' state appears at lower gain $\Gamma$ compared to ``symmetric-II'' state and can be seen as a kind of the counterpart of the antisymmetric state. The absolute value of the field in the middle resonator is higher in ``symmetric-I'' than in ``symmetric-II'' state for the same gain $\Gamma$. For the latter state the ratio of the field intensity in the middle resonator to the field intensity in $B$ (or C) resonator goes to zero with the increase of the gain. This explains why the bifurcation curve of `symmetric-II'' state approaches the bifurcation curve of the antisymmetric state at large $\Gamma$, see Fig.\ref{fig1}(b).

The amplitudes of the fields can be found analytically for symmetric-I state 
\begin{subequations}
	\begin{equation}
    \omega = \alpha \dfrac{\Gamma}{\beta} + \delta + \dfrac{2\sigma^2}{\beta(\gamma^2 + \omega^2)}(\beta\omega-\alpha\gamma),
	\end{equation}
	
	\begin{equation}
    |B_0|^2 = \dfrac{\Gamma}{\beta} - \dfrac{2\sigma^2\gamma}{\beta(\gamma^2+\omega^2)},
	\end{equation}
	
	\begin{equation}
    A_0 = 2\sigma\dfrac{\omega + i\gamma}{\gamma^2 + \omega^2}|B_0|.
	\end{equation}
\end{subequations}
The numerical results of the stability analysis for `symmetric-I'' state are summarized in Fig.\ref{fig2}(c). So, for $\delta \lesssim \delta_{cr5} \approx - 6.1$ `symmetric-I'' state is stable until the gain $\Gamma$ is below a critical value, but the state looses its stability at higher gain through subcritical Hopf bifurcation. Such instability results in switching of the system to the antisymmetric state. For the intermediate range of detunings $ \delta_{cr5} \lesssim \delta \lesssim \delta_{cr6} \approx 0.12$ the `symmetric-I'' state becomes unstable at gains close to the excitation threshold too. This instability is also provided by subcritical Hopf bifurcation; its development leads to the formation of the anti-symmetric state. Finally, at $\delta_{cr6}$ the stability region of ``symmetric-I'' state vanishes and then the states are unstable within the whole range of existence $\delta<\delta_{cr7}\approx 9.91$.

The symmetric-II states were found numerically and it was found that these state can also be stable for $\delta>\delta_{cr8}=-1.8$, the narrow tongue of stability is clearly seen in Fig.\ref{fig2}(e). Decreasing linear gain from the region of stable solutions leads to subcritical Hopf bifurcation and unstable symmetric state relaxes to the anti-symmetric state as a results of the instability development. Increase of the gain leads to spontaneous symmetry breaking bifurcation and appearing of the hybrid states (double degenerated) of the second kind. The bifurcation curve for hybrid-II states is shown in Fig.\ref{fig1} by the brown dashed curve. Such hybrid states are stable only when bifurcate from the stable symmetric state and have very narrow stability range. Both symmetric-II and hybrid-II states have small stability area in the parameter space. On top of it these stable states have small basins of attractions. Therefore these states are less interesting from the physical point of view because it seems to be a very hard problem to observe them experimentally.

In next Section we report the results of numerical simulations on the formation of the different nontrivial states from weak noise and the switching between the states.

\section{Direct numerical simulations of the trimer dynamics}

The existence of a stable state means that this state can be observed in experiments. However, the basin of attraction of some states is so small that they can form only from the initial conditions very close to the exact stationary solution. It can happen to be difficult to create the necessary initial conditions in the experiment and thus it can be a challenging task to observe these states. Other states can form of the weak noise taken as initial conditions and one can assume that these states are much easier to observe in experiments. In this section we study the formation of the states from the weak noise and the switching of the states caused by the change of the linear gain.      

For this purpose we perform numerical simulation of (\ref{eq:trimer}) with  $\Gamma$ varying in time. We set the detuning to be $\delta = -2.4$ because this set of parameters allows to observe most interesting dynamics, in particular most various dynamics of the hybrid states.

\begin{figure}[t]
\centering
\includegraphics[width=\linewidth]{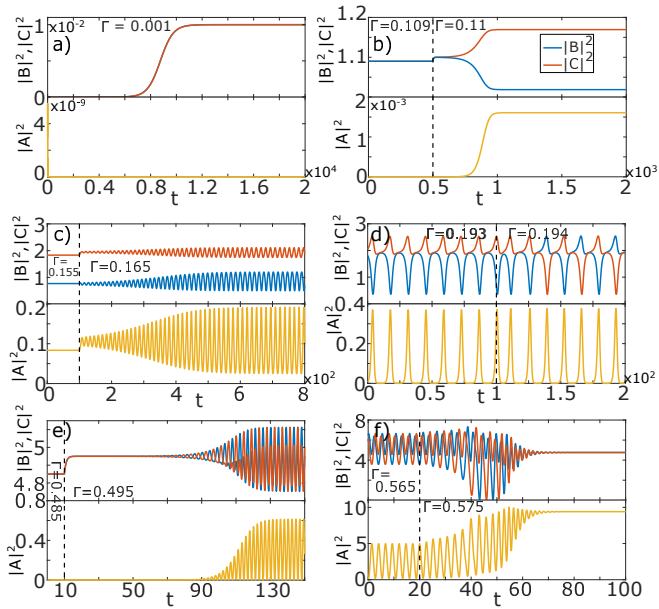}
\caption{(a) Evolution in time of the antisymmetric stationary state from the weak initial noise; (b) Switching of the system to hybrid state by increasing of gain from $\Gamma = 0.109$ to $\Gamma = 0.11$ at $t=500$; (c) Appearing of the oscillating state from the hybrid state because of the increase of linear gain to $\Gamma=0.165$ at $t=1000$; (d) Switching between two types of oscillating states at $\Gamma=0.194$; (e) Formation of oscillating state from the antisymmetric state; (f) Switching of oscillating state into the symmetric state. At each panel begining of time interval is shifted to $t = 0$ for convenience of perception. Parameters are the same as in Fig.\ref{fig1}, but $\delta = -2.4$}
\label{num}
\end{figure}

We start our numerical simulation with near-zero value of linear gain ($\Gamma = 0.001$) and take the initial conditions in the form of weak noise. As it can be clearly seen from Fig.\ref{num}(a) the amplitudes of the fields in both B and C resonators start growing exponentially and the saturation of the growth results in the formation of the anti-symmetric state. In the lower part of panel (a) one can clearly seen that the middle resonator A is not excited at all which is a signature of the anti-symmetric state. 
It complies well with the analysis of stationary states summarized in Fig.\ref{fig1} where one can see that at small values of the gain the anti-symmetric state is the only one possible stationary solution. 

Now we take the anti-symmetric solution already formed from the noise and start increasing the gain. When the gain $\Gamma$ exceed the threshold of the symmetry breaking bifurcation then the system switches to a hybrid state with broken symmetry, see  Fig.\ref{num}(b) showing this switching. One can see that in the final state the intensities of the filed in resonator B and C are different and the resonator A is excited.  Increase of $\Gamma$ makes the intensity difference between two active resonators larger and the intensity of passive resonator A increases too.

If the linear gain $\Gamma$ exceed the threshold value when the hybrid-I state gets destabilised via  supercritical Hopf (see the section ``Hybrid states''), then the system switches from the hybrid to oscillating state, see Fig.\ref{num}(c). Since the oscillating state bifurcates from the stationary state with broken symmetry, the amplitudes of oscillations are different for both active resonators. With the increase of $\Gamma$ the amplitude of the oscillations growth and at some point the oscillating state switches to another oscillating state, see Fig.\ref{num}(d). In the new oscillating state the amplitudes of the fields in resonator B and C become equal. 

%With bigger values of $\Gamma$ oscillating state switches to another dynamical state, see Fig.\ref{num}(d). This state is also oscillating, but the amplitudes of oscillation are identical for both lasers. 

It is interesting that increasing the gain further one can switch the system back to the oscillating state with different amplitudes in resonators B and C and then to a non-oscillating hybrid-I states. It was also observed in numerical simulations that finally the hybrid states restore the symmetry and transform to the anti-symmetric state.  

% With further increasing of $\Gamma$ the oscillating state from right part of Fig.\ref{num}(d) switches again to the oscillating state with different intensities of active resonators. Qualitatively, such plot is similar to the Fig.\ref{num}(d) with reversed time, that is why there is no additional panel for such switching. With larger linear gain the system through supercritical Hopf bifurcation switches to the stationary state with broken symmetry again and after it to the antisymmetric state. These processes are similar to transitions demonstrated in Fig.\ref{num}(c,b) with reversed time.

Now let us study destabilization of the anti-symmetric states by the Hopf bifurcation. When the linear gain increases the threshold value an oscillations of the field intensities start growing destabilizing the system. As a result of this instability a new oscillating state forms, see Fig.\ref{num}(e). For this oscillation state the fields in the resonators B and C oscillate in anti-phase. Further increase of the gain leads to destabilization of the oscillating state and the system switches to the symmetric state, see Fig.\ref{num}(f). This state is characterized by high intensity of the filed in middle resonator A and is, of course, dynamically stable. We identify the symmetric state as symmetric-I state described above.

We did not manage to observe hybrid-II and symmetric-II states in numerical simulations what can be explained by small basins of attraction of these states. However, it is important observation that the formation of the anti-symmetric, hybrid-I and symmetric-I states can be easily formed in the system by the appropriate manipulation by the linear gain $\Gamma$. At least three different oscillating states can also be observed in the system.

\section{CONCLUSION}
In this paper we have considered the dynamics of the system of two nonlinear active resonators coupled through a linear passive one. The stationary states of such trimer are investigated and classified in terms of symmetry in ``antisymmetric'', ``symmetric'', and ``hybrid'' states. It is demonstrated, that with antisymmetric state the radiation is locked inside the active resonators because of the destructive interference in the linear one. This state is in certain sence equivalent to the effect of bound state in the continuum (or BIC) when the radiative losses are completely compensated by the interference effect. The symmetric state, on the contrary, excites the middle resonator what can be seen as an analogue of constructive interference increasing the radiative losses in BIC-like systems.

Both antisymmetric and symmetric states are characterized by equal absolute field amplitudes in active resonators. In turn, the hybrid stationary states appear due to a symmetry breaking bifurcation and thus they differs from the symmetric and anti-symmetric states having nonidentical field amplitudes in the nonlinear resonators B and C. These hybrid states are characterized by non-zero field in linear resonator.

The dynamical stability and bifurcations of all the stationary states are analyzed in detail. It is demonstrated, that the state with broken symmetry can bifurcate both from the antisymmetric and the symmetric stationary states. However, the hybrid state bifurcating from the symmetric state has poor stability which makes its observation hard to do. The hybrid state bifurcating from the antisymmetric state is stable but can lose its stability through Hopf bifurcation resulting in formation of oscillating quasi-periodic state.

The formation of the anti-symmetric state from weak noise is demonstrated by direct numerical simulations. It is also shown that changing the gain one can transform the anti-symmetric states to the hybrid state. By further manipulations with the gain oscillating and the symmetric states can be obtained in the system. 

Thus we can conclude the considered system consisting of three interacting resonators allows to observe different states and switch them in a controllable way. This can be of interest from the point of view of the design of the dynamically re-configurable micro-lasers and find applications in the field of coherent light generation and optical simulations.    

\begin{acknowledgments}
The authors acknowledge the financial support provided by Russian Fund for Basic Research (Grant ``Aspiranty'' No. 20-32-90227).
\end{acknowledgments}
\bibliography{sample}% Produces the bibliography via BibTeX.

\end{document}